# Title: Chiral ground states of ferroelectric liquid crystals

**Authors:** Priyanka Kumari[1,2], Bijaya Basnet[1,2], Maxim O. Lavrentovich[3], Oleg D. Lavrentovich[1,2,4]*

**Affiliations:**

[1]Advanced Materials and Liquid Crystal Institute, Kent State University; Kent, OH 44242, USA.

[2]Materials Science Graduate Program, Kent State University; Kent, OH 44242, USA.

[3]Department of Earth, Environment, and Physics, Worcester State University; Worcester, MA 01602, USA.

[4]Department of Physics, Kent State University; Kent, OH 44242, USA.

*Corresponding author. Email: olavrent@kent.edu

**Abstract:** Ferroelectric nematic liquid crystals are formed by achiral molecules with large dipole moments. Its three-dimensional orientational order is universally described as unidirectionally polar. We demonstrate that the ground state of ferroelectric nematic unconstrained by externally imposed alignment directions is chiral, with left- and right-hand twists of polarization. Although the helicoidal deformations and defect walls separating domains of opposite handedness increase the elastic energy, the twists reduce the electrostatic energy and become weaker when the material is doped with ions. The study shows that the polar orientational order of molecules could trigger chirality in the soft matter with no chemically induced chiral centers.

**One-sentence Summary:** Ferroelectric nematic liquid crystals form macroscopically chiral structures even when the constituent molecules are not chiral.

**Main Text:** The emergence of chirality in living organisms and soft matter in general is a fascinating field of current research. Usually, macroscopic chirality requires chemical chirality of constituent molecules. The well-known examples are nematic and cholesteric liquid crystals. A nematic (N) formed by elongated achiral molecules shows a uniaxial 3D orientational order. This orientation order, described by an apolar unit vector $\hat{\mathbf{n}} \equiv -\hat{\mathbf{n}}$, called the director, does not distinguish between the heads and tails of the molecules even when these are chemically different (*1*). When the molecules are chiral, a cholesteric (N*) forms (*1*). Locally, over a nanometer scale, cholesteric is similar to a nematic, as the molecules appear to align parallel to each other. However, globally, the N* director follows a helicoid, rotating in space around an axis that is perpendicular to the long axes of molecules. The helicoidal pitch $\mathcal{P}$, typically longer than ~100 nm, is much larger than the molecular scale of 1 nm, which reflects the relative weakness of the chiral molecular interactions.

A polar version of a nematic has been synthesized and characterized (*2-8*). This liquid crystal, identified as a 3D uniaxial ferroelectric nematic ($N_F$) (*6*), is formed by achiral rod-like molecules with large permanent longitudinal electric dipoles, which align parallel to each other, producing spontaneous macroscopic polarization **P** along the molecular director $\hat{\mathbf{n}}$. Most studies to date focus on $N_F$ samples with well-defined surface alignment of **P** imposed by unidirectional rubbing of the polymer coatings at the glass plates confining the material. Since liquid crystals transmit torques (*1, 9*), the bulk structure is determined by the surface interactions. The equilibrium state of a flat $N_F$ slab is thus either a uniaxial monocrystal (for parallel assembly of unidirectionally rubbed plates) or $\pi$-twisted (antiparallel assembly) (*6, 7, 10-12*). In this work, we design $N_F$ slabs that are free of the surface-imposed torques: the vector **P** is unidirectionally aligned by rubbing at one substrate but is free to rotate at the opposite surface. The equilibrium state of these $N_F$ cells is chiral rather than unidirectionally polar: the films split into domains with alternating left- and right-hand twists of **P**. No such chirality is observed when the material is heated from the $N_F$ phase into the paraelectric N. The twists weaken when the material is doped with ionic additives.

We explore two well-studied $N_F$ materials, Fig.1: (i) DIO, reported by Nishikawa et al. (*3*), synthesized in the laboratory as described in (*12*), and RM734, synthesized by Mandle et al. (*2*), and purchased from Instec, Inc. On cooling from the isotropic (I) phase, the phase sequence of DIO is $I-174°C-N-82°C-SmZ_A-66°C-N_F-34°C-Crystal$, where $SmZ_A$ is an antiferroelectric smectic with a partial splay (*13*). The phase sequence of RM734 is $I-188°C-N-133°C-N_F-84°C-Crystal$.

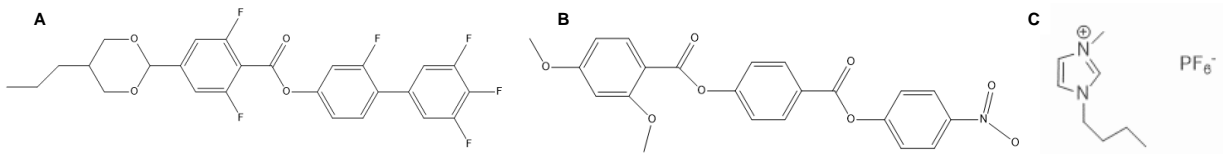

**Fig.1. Chemical structure of the explored materials. (A, B)** Ferroelectric nematic liquid crystals DIO and RM734, respectively. **(C)** Ionic liquid BMIM-$PF_6^-$.

The bottom substrates of all cells are glass plates with a layer of polyimide PI2555 rubbed unidirectionally along the direction **R**, written in Cartesian coordinates as $\mathbf{R}(z = -h/2) = (0, -1, 0)$; the $xy$ plane is parallel to the bounding substrates, and the coordinate $z = -h/2$ is assigned to the location of the rubbed plate; $h$ is the cell thickness. DIO aligns the polarization antiparallel to the buffing direction, $\mathbf{P} \uparrow\downarrow \mathbf{R}$ (*12*), while $\mathbf{P} \downarrow\downarrow \mathbf{R}$ in the case of RM734 (*6*). In both cases, the "pretilt" angle between **P** and the substrate is negligibly small, to avoid a strong surface charge (*11, 12, 14, 15*); this angle is estimated as $10^{-7}$ (*15*). The opposite surface, $z = h/2$, is tangentially degenerate, allowing **P** to point along any direction within the $xy$ plane, remaining in this plane to avoid surface charges. We use three different top surfaces, all setting azimuthally degenerate in-plane alignment of **P**: (a) a glass plate coated with polystyrene (PS) (*16, 17*); PS is one of a few polymer coatings that is free of anchoring memory described by N. A. Clark (*18*); (b) a glass plate coated with a thin layer of glycerin (*17, 19*); (c) free surface (air) (*17*). The two plates in (a) and (b) are separated by spacers that fix $h$ in the range 2-10 μm, which is constant within ±0.2 μm. In order to separate the electrostatic mechanism of chirality from the so-called geometrical anchoring (*20*), we also explored wedge samples with a very small dihedral angle $\alpha = 0.3°$, in which **R** is perpendicular to the thickness gradient to avoid splay deformations, Fig.2A-E.

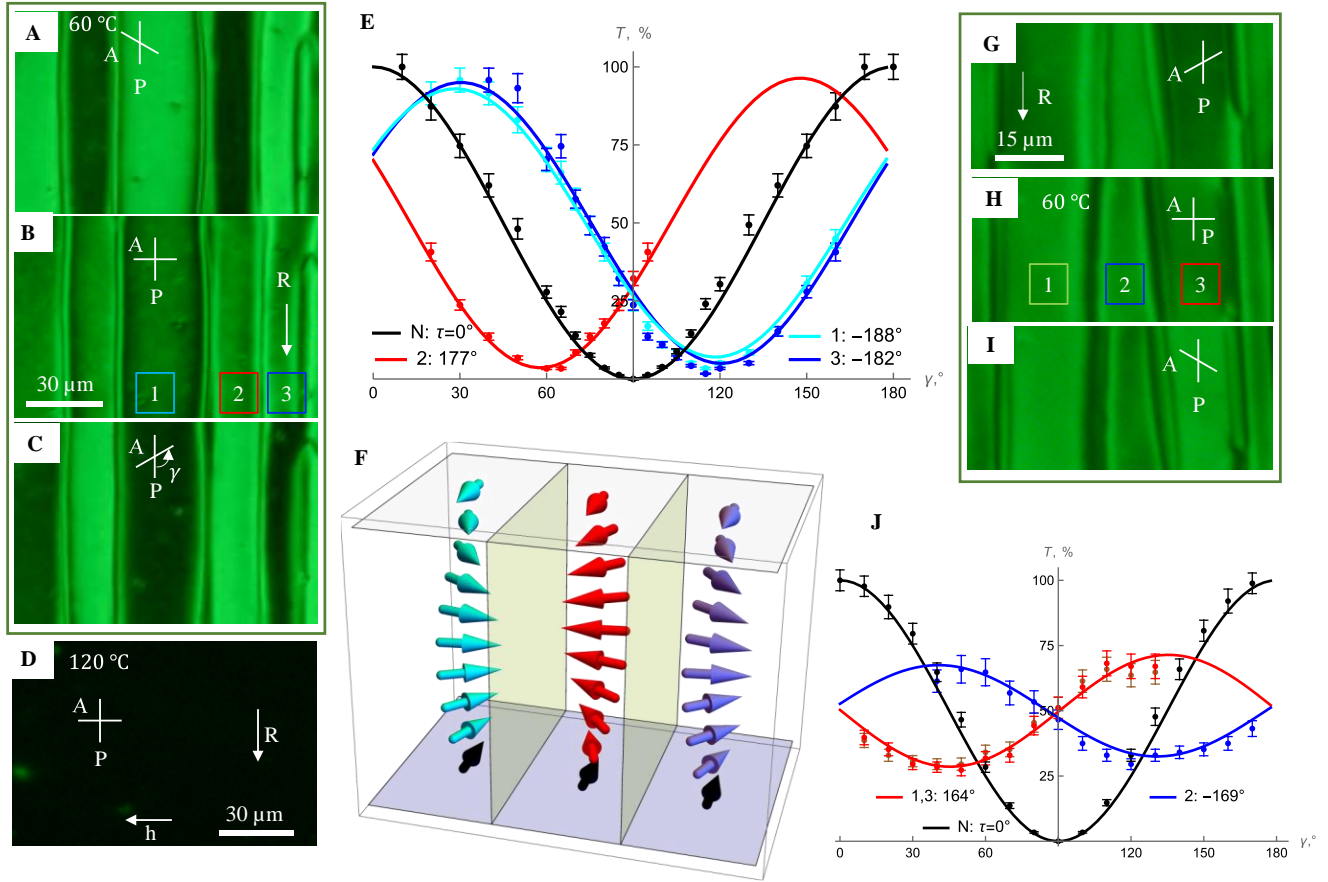

**Fig.2. Chiral ground state of DIO N$_F$ cells confined between PI2555 and PS.** (**A, C**) Polarizing microscope textures of a wedge cell ($h = 2$ μm) for uncrossed polarizer and analyzer; (**B**) the same domains, crossed polarizer and analyzer; (**D**) extinction texture of the N phase in the same cell; the horizontal white arrow shows the direction of thickness increase; (**E**) transmitted light intensity $T$ through N$_F$ domains 1, 2, 3 in (B) and also for the N phase as a function of the angle $\gamma$ between the polarizer and analyzer; solid curves are numerical fits with the twist angle $\tau$ as the fitting parameter; (**F**) left- and right-handed twists of **P** in neighboring N$_F$ domains separated by domain walls; (**G, H, I**) textures of a flat cell ($h = 3$ μm); (**J**) transmitted light intensity $T$ vs. $\gamma$ for N$_F$ domains 1, 2, 3 in (H).

The cells are explored between two linear polarizers. The plane of polarization of the bottom (entry) polarizer is always along the $y$-axis, Fig.2, parallel to the polarization **P** at the PI2555 coated plate, $z = -h/2$. The top analyzer can be rotated; the angle $\gamma$ between the analyzer and polarizer varies from 0 to 180°. All cells show qualitatively similar behavior, as demonstrated for a PI2555/DIO/PS wedge cell of type (a) with $h = 2$ μm in Fig.2A-E, flat PI2555/DIO/PS cell with $h = 3$ μm in Fig.2G-J, flat cells of types (a) and (b) with DIO and RM734 in Figs.S1, S2, and S3, cell of type (c) in Fig.S4, cell with added colloidal spheres that create axial distortions with the symmetry axis along the local **P** in Fig.S5. Below, we discuss the PI2555/DIO/PS cells shown in Fig.2.

In the N phase, the director adopts the orientation imposed by the bottom plate's rubbing direction **R**, so that $\hat{\mathbf{n}} = (0, \pm1, 0) = \text{const}$ everywhere in the bulk. When viewed between crossed polarizers, $\gamma = 90°$, the texture is dark, as the propagating extraordinary beam is extinguished by the analyzer. The light transmittance at this setting is set as zero, $T = 0\%$, Fig.2D. As $\gamma$ deviates from 90°, the texture brightens, following the dependence $T(\gamma) \propto \cos^2\gamma$, Figs.2E, J. We do not observe chiral segregation reported in similarly treated N cells of DIO [7].

When the sample is cooled down to the $N_F$ phase, the homogeneous texture transforms dramatically into a pattern of elongated domains, Figs.2A-C, G-I. These domains transmit a significant amount of light when $\gamma = 90°$; $T \approx 50\%$ in Figs.2H,J. Smaller levels of light transmission occur when the analyzer rotates clockwise (CW) or counterclockwise (CCW). The contrast of the neighboring domains alternates. For example, in Figs.2A,C, when $\gamma \approx 60°$, domain 2 is dark, while domains 1 and 3 are bright, Fig.2A. Conversely, when $\gamma \approx 120°$, domains 1 and 3 are dark and domain 2 is bright, Fig.2C. The light transmittance by $N_F$ domains as a functions of $\gamma$ is markedly different from the dependence $T(\gamma) \propto \cos^2\gamma$ in the N phase, Figs.2E,J. The data provide evidence that the optical axis $\hat{\mathbf{n}}$ and **P** in the $N_F$ domains twist along the axis $z$ perpendicular to the cell, as schematized in Fig.2F. The samples exhibit approximately equal amounts of left- and right-handed helical twists, indicating that the effect cannot be attributed to the presence of chiral admixtures.

The light beam impinging onto the bottom plate of an $N_F$ cell is polarized along **R**, $\hat{\mathbf{n}}$, and **P**. In the so-called waveguiding (or Mauguin) regime, the light propagating along the helical axis of a twisted liquid crystal remains linearly polarized; this polarization rotates with $\hat{\mathbf{n}}$ and **P**. However, the waveguiding regime requires the so-called Mauguin number to be very large, $Mau = \frac{\pi(n_e - n_o)d}{\lambda\tau} \gg 1$; here $\lambda$ is the wavelength of the probing light ($\lambda = 532$ nm in all our experiments); $\tau$ is the angle by which $\hat{\mathbf{n}}$ and **P** twist between the bottom and the top plates; $n_e$ and $n_o$ are the extraordinary and ordinary refractive indices, respectively, Fig.S6. In the experiments above, $Mau$ is less than 10, thus the propagating light is elliptically polarized, as evidenced by the significant transmission of light at any $\gamma$ in Fig. 2J, where the smallest transmission is very significant, $T = 30\%$. For DIO, $n_e - n_o \approx 0.19$, Fig.S6. So, even if one assumes that the locations of transmission minima $\gamma \approx 60°, 120°$ in the cell with $h = 2$ μm, Fig.2E, correspond to the weakest possible twist, $\tau = \pm30°$, $Mau$ is still only about 4. For $Mau \leq 10$, the polarization of transmitted light generally changes from linear to elliptical so that the analyzer cannot reduce the transmitted intensity to zero at any $\gamma$. For elliptically polarized light, $\tau$ could be estimated by numerical simulations based on Jones matrices [21, 22], see Methods. Briefly, the cell is divided into $N$ slabs of thickness $\delta d = h/N$, thin enough to be approximated as uniform nematic-like layers with a well-defined uniaxial optic axis. This axis then rotates between adjacent slabs by an angle $\tau/N$. The director remains in the $xy$ plane of the sample, so that each slab introduces an optical retardance $\Gamma = \frac{2\pi h}{\lambda N}(n_e - n_o)$. The director twist is assumed to be a linear function of the coordinate $z$, as in the well-studied twisted N cells [21]. This is a strong assumption since the neighboring left- and right-twisted $N_F$ domains are separated by domain walls, in which **P** acquires additional deformations to avoid head-to head and tail-to tail strongly charged arrangements,

Fig.2F. Since liquid crystals transmit torques, the deformations within the domain walls propagate along all three spatial directions, which complicates the structure. The Jones matrices of all $\delta d$-slabs are multiplied together to yield the overall optical response of the cell, which is independent of $N$ for sufficiently large $N$.

Numerical simulations demonstrate that for DIO $N_F$ cells with $h = 2$ μm, the twist angle $\tau$ between the bottom and top surfaces is about 180°. For example, in Fig.2E, $\tau_1 = -188°$, $\tau_2 = 177°$, $\tau_3 = -182°$, where the subscripts correspond to the numeration of domains in Fig.2B. For a cell in Figs. 2G-J, $\tau_1 = \tau_3 = 164°$, and $\tau_2 = -169°$. The fitted $\tau$'s are accurate within $\pm 10\%$, Fig.S7. These $\tau$ values suggest that the helical pitch is $\mathcal{P} \approx 2h \approx (3.8 - 4.1)$ μm in the $h = 2$ μm cell and $\mathcal{P} \approx (6.4 - 6.6)$ μm in the $h = 3$ μm cell. The observed increase of $\mathcal{P}$ with $h$ suggests that the twisted states are rooted in the spatial confinement of the samples and the balance of elastic energy cost and electrostatic energy gain. In some cases, numerical simulations produce multiple values that fit the experiments reasonably well. However, for the $h = 3$ μm cell, Fig.2J, there are no fitted $\tau$'s that agree with the experimental data other than the reported above in the entire range from $\mathcal{P} \approx (6.4 - 6.6)$ μm to $\mathcal{P} \approx 0.4$ μm. A pitch $\mathcal{P} \approx (0.4 - 0.5)$ μm should cause selective reflection of light in the visible part of the spectrum, but we do not observe such an effect. Given the limitation of the numerical simulations caused by the assumption of linear twist, more refined experiments are needed to measure the pitch accurately and to establish its dependence on the cell size, presence of ions, temperature, etc.

The alternating left and right twists in the $N_F$ domain structures are confirmed by the observations in circularly polarized light, Fig.S8.

The twists diminish in $N_F$ cells doped with the ionic fluid 1-Butyl-3-methylimidazolium hexafluorophosphate (BMIM-PF₆). At weight concentration 0.5 wt% of BMIM-PF₆, which corresponds to $1.5 \times 10^{25}$ m$^{-3}$ ions when the ionic fluid is fully ionized, the $N_F$ phase preserves its ferroelectric ordering, as evidenced by the polar response to a direct current (dc) electric field and by the recent studies of the Colorado group (23). However, the doped $N_F$ confined between a rubbed PI2555 plate and a PS plate while preserving some twisted domains with bright textures between crossed polarizers, Fig.3A, also shows areas where the texture is extinct, with a vanishingly small twist $\tau$, Fig.3B. Similar reduction of twists is observed when the doped DIO is placed between PI2555 and a layer of glycerin, Fig.S9.

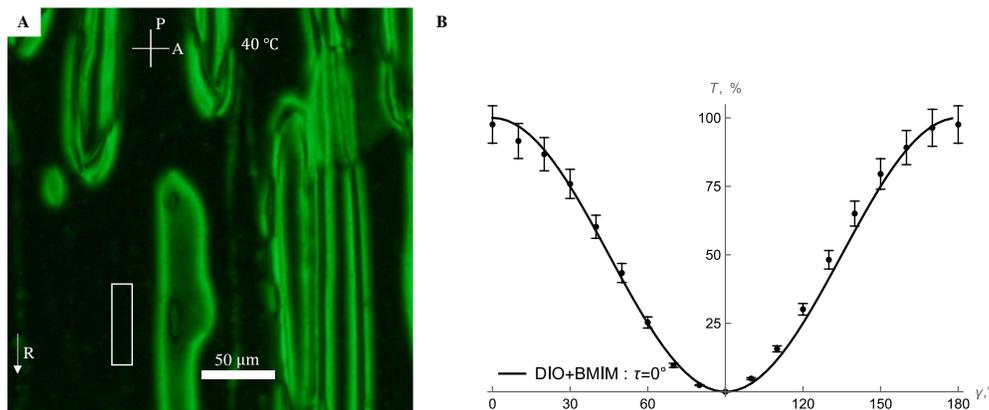

**Fig. 3. Partially unwound ground state of $N_F$ flat cell of DIO doped with 0.5 wt. % of the ionic fluid.** (**A**) Polarizing microscope texture for crossed polarizers; bright domains with twist alternate with dark regions of a vanishing twist; (**B**) transmitted light intensity vs. $\gamma$ measured within the white rectangle in part (A) illustrating a vanishing twist angle. The $N_F$ is confined

between a rubbed PI2555 plate and a PS plate, which sets azimuthally degenerate tangential anchoring. Cell gap $h = 7$ μm.

Finally, a dc electric field **E** of a small amplitude 20 kV/m applied along the direction of **P** at the PI2555 plate realigns **P** parallel to itself everywhere and thus removes the twisted domains, Fig.S10; the textures become homogeneous and extinct in crossed polarizers, Fig.S10B,C. When the field is switched off, the chiral domains reappear, Fig.S10D-G. The twists restore slowly, within 30-60 min, since there is no surface anchoring torque to speed the relaxation up and since the process involves the diffusive redistribution of ions once the field is switched off.

**Discussion**. Ferroelectric nematics are a fluid version of the better-explored crystal ferroelectrics. Crystal ferroelectrics consist of domains in which the polarization **P** is along one of the crystallographic axes. The domain structure lowers the electrostatic energy. In most cases, the crystallographic axes inhibit a continuous rotation of **P** (24).

In the ferroelectric nematic fluid, there are no crystallographic axes, and the neighboring directions of **P** might reorient by an infinitesimally small angle. The observed left-and right-twisted domains in the ground state of cells with no surface-imposed torques are caused by the material's tendency to reduce the electrostatic energy by replacing the three-dimensional unidirectional polar orientation of **P** with left and right helicoidal twists, **P** $\propto [\cos(2\pi z/\mathcal{P}), \pm\sin(2\pi z/\mathcal{P}), 0]$. This mechanism of electrostatic energy reduction has been theoretically foreseen by Khachaturyan in 1974 (25).

Khachaturyan's model considers a single domain with a linear twist, **P** $\propto [\cos(2\pi z/\mathcal{P}), \sin(2\pi z/\mathcal{P})]$, which makes the structure a polar analog of a conventional cholesteric (25). The predicted pitch is $\mathcal{P} = 2\pi(\xi^2 r)^{1/3}$, where $r$ is the typical radius of the sample measured in the $xy$ plane perpendicular to the twist axis, $\xi = 2\sqrt{\frac{\varepsilon_a \varepsilon_0 K_{22}}{P^2}}$, $\varepsilon_a > 0$ is the dielectric anisotropy, $\varepsilon_0$ is the electric constant, and $K_{22}$ is the twist elastic constant (25). The pitch $\mathcal{P}$ depends on the size of the sample, in drastic contrast to the behavior of cholesterics, in which $\mathcal{P}$ is an intrinsic property defined by molecular interactions and only slightly affected by spatial confinement. For $P = 4.4 \times 10^{-2} \frac{C}{m^2}$, $K_{22} = 5$ pN, assuming that $\varepsilon_a$ could range in the wide range from 10 to $10^4$, one estimates $\xi = (1 - 30)$ nm. If the width of domain, $r = 20 - 100$ μm, is the measure of the sample size, then one expects $\mathcal{P} = (0.2 - 3)$ μm. The range $\mathcal{P} = (0.2 - 0.5)$ μm should be ruled out in our experiments since we do not observe selective reflection of light. The estimate assumes no screening by ionic impurities. These are expected to reduce the effective |**P**| and thus increase $\mathcal{P}$. Experiments with DIO doped with the ionic fluid, Fig.3, confirm that the pitch increases as the polarization is screened by ions. Therefore, the experimental estimates $\mathcal{P} = (4 - 7)$ μm appear to be close to the theoretical expectation if one accounts for the inevitable presence of ionic impurities. Divergence of $\mathcal{P}$ by Debye screening has been qualitatively predicted by Khachaturyan (25); unfortunately, the corresponding formula is dimensionally incorrect.

One should also consider a second potential mechanism of the chiral ground state of the ferroelectric fluids, rooted in the so-called geometrical anchoring (20). The twist could be caused by the wedge geometry of confinement in which one plate sets **R** *along the thickness gradient*, and the other plate yields an azimuthally degenerate tangential anchoring. If the director (or **P**) in the bulk follows **R**, such a cell features splay. The splay energy could be reduced by twisting the director (and **P**) towards the direction perpendicular to the thickness gradient. However, the possibility of geometrical anchoring triggering twisted domains is excluded by the design of the wedge cell in Fig.2A-E, in which the rubbing direction **R** is *parallel* to the edge. Such a geometry with **P** parallel to the edge allows the system to be free of any deformations and carry no elastic energy and yet the twists appear to reduce the electrostatic energy. Moreover, the appearance of twist in a wedge cell, Fig.2A-C, necessarily implies a small splay since the two plates are tilted with respect to each other (*26*). Therefore, the design of wedge cells hinders the electrostatics-triggered twists rather than facilitates them.

In flat cells, one can consider the possibility that the cell gap varies across the cell extension in the $xy$ plane, which is about 1 cm. For the typical gap variation $\delta h = 0.2$ μm, the unintentional dihedral angle is $\alpha \approx 0.2 \times 10^{-4}$. The free elastic energy per unit area of such a film with splay replaced with a twist $\tau < 90°$ writes (*20*)

$$f = \frac{K_{11}}{2d}[\arcsin(\arcsin\alpha\cos\tau)]^2 + \frac{K_{22}}{2d}[\arctan(\tan\alpha/\cos\tau)]^2,$$

For small $\alpha$, the equilibrium twist angle $\tau$ is defined by the equation

$$\alpha^2\frac{K_{11}}{K_{22}} = \frac{2\tau}{\sin 2\tau}.$$

The factor $\frac{2\tau}{\sin 2\tau}$ is always larger than 1. Therefore, the twist could occur for $\alpha \approx 0.2 \times 10^{-4}$ only when the ratio of the two elastic constants is truly enormous, $\frac{K_{11}}{K_{22}} \sim 10^9 - 10^{10}$, which does not appear to be realistic since one does observe some splay regions extending over a few micrometers in the $N_F$ textures (*6, 12, 17*). Furthermore, the variations of $\delta h$ are random within the nominally flat cells, while the observed twist domains show quasi-one-dimensional periodicity in all studied cells, which is especially clear when the field of view is much larger than the width of the domains, Fig.S8. Surface imperfection, details of the domain walls' structure, and nonuniform distribution of ions might contribute to variations of the domains' width, Figs.2, S1-S3, S5, S8-S10. Importantly, the geometrical anchoring could not explain twists larger than 90°, and the strong light transmission in Figs.2G-J demonstrates that $\tau > 90°$. Therefore, the geometrical anchoring is not the reason for the observed chiral states.

To conclude, we demonstrated that the ground state of a ferroelectric nematic liquid crystal unconstrained by the surface-imposed torques is chiral. The sandwich-like cells split into an array of elongated domains with alternating left- and right-handed helical twists. The twists weaken when the ferroelectric liquid is doped with an ionic fluid. The domains are also removed by applying a dc electric field; once the field is switched off, the domains reappear. These chiral states are difficult to observe in conventional cells in which the two bounding plates are rubbed unidirectionally. For example, it is well known that in thin cells of a cholesteric in which the gap is smaller than some critical value $h_c$ defined by a balance of the elastic and azimuthal anchoring torques as $h_c = 2\pi K_{22}/W$ (*27*), where $W$ is the surface anchoring coefficient (assumed to contain only a quadrupolar contribution and no in-plane polarity term), the cholesteric unwinds from a $\pi$-twist to a no-twist monocrystal state; for the typical $K_{22} = 5$ pN, $W \sim 10^{-5}$ J/m$^2$ (*12*), one finds $h_c \approx 3$ μm, which is close to the typical cell thicknesses in the experiments on ferroelectric fluids performed so far.

The observed chiral states should be distinguished from the chiral state of a cholesteric liquid crystal, which is caused by the chemically-introduced molecular chirality (*1*): there are no chiral molecules in the explored materials. These should also be distinguished from the twisted states observed in confined samples of an achiral N with curved interfaces (*19, 26, 28*), in which twists replace confinement-induced splay and bend when $K_{22}$ is the smallest of the bulk elastic constants. In our experiments with flat and wedge cells, the spatial confinement allows the bulk structure to remain uniformly aligned with zero elastic and anchoring energies, and yet the system prefers to split into right- and left-twisted periodic domains separated by domain walls. Although these twisted states increase the elastic energy, they reduce the electrostatic energy associated with a unidirectional polar orientation of the spontaneous electric polarization **P**. A proper abbreviation for the described ground state of the liquid ferroelectric could be $N_{F*}$, where the subscript "*" refers to the macroscopic chirality of the chemically achiral material. A ferroelectric liquid formed by chiral molecules could be abbreviated $N_{F*}^*$ where the superscript "*" is widely accepted as a reference to chemical chirality.

The emergence of macroscopic chirality in the $N_F$ is associated with the frustration between the tendency to form a locally polar uniaxial order and the increase of the electrostatic energy of the macroscopic volume in the presence of such a polar order. The observation of chiral states poses numerous new questions about the properties of $N_F$, such as dependency of the twist pitch on the sample size, ionic content, presence of chiral impurities, surface interactions, temperature, electromagnetic fields, etc. It is also intriguing to explore whether the polarity-triggered chirality could be found in other soft materials, including those of a biological origin.

**Acknowledgments:** The authors thank Dr. Hari Krishna Bisoyi and the Organic Synthesis Facility at the AMLCI for the synthesis and purification of DIO and RM734.

**Funding:** This work is supported by NSF grant ECCS-2122399 (O.D.L.).

**Author contributions:**

Experiments: P.K. and B.B.

Experimental data analysis: P.K., B.B. and O.D.L.

Numerical analysis: M.O.L. and O.D.L.

Idea and design of experiment: O.D.L.

Writing: O.D.L. with input from all coauthors.

**Competing interests:** The authors declare no competing interests.

**Data and materials availability:** The authors declare that the data supporting the findings of this study are available in the main text or the supplementary materials.

## Supplementary Materials
Materials and Methods

Figs. S1 to S10

<div align="center">

Supplementary Materials for

**Chiral ground states of ferroelectric liquid crystals**

</div>

<div align="center">

Priyanka Kumari[1,2], Bijaya Basnet[1,2], Maxim O. Lavrentovich[3], Oleg D. Lavrentovich[1,2,4]*

Corresponding author: olavrent@kent.edu

</div>

**The PDF file includes:**

Materials and Methods

Figs. S1 to S10

## Materials and Methods

DIO. The name "DIO" was introduced by Nishikawa et al., who were the first to synthesize the material (*3*). The dioxane ring of DIO molecules could be in *trans*- or *cis*- conformations. The *cis*-isomer is not mesomorphic, and its presence significantly decreases the transition temperatures (*8*). For example, the I-N transition temperature in the *cis:trans* = 10:90 composition is 150° C, which is by 24° C lower than the temperature 174° C of the I-N transition of a pure *trans* composition. The transition temperatures of the DIO synthesized in our laboratory (*17*) are close (within 2° C) to the ones reported by Nishikawa et al. (*3*) for the *trans* composition. Therefore, the DIO studied in this work comprises the *trans*-isomers.

RM734. The name "RM734" was introduced by Mandle et al., who were the first to synthesize the material (*2*). The material was purchased from Instec, Inc. (purity better than 99%) and purified by silica gel chromatography and recrystallization in ethanol.

Ionic liquid. 1-Butyl-3-methylimidazolium hexafluorophosphate (BMIM-PF$_6$), a room temperature ionic fluid of density (1.371 - 1.375) g/cm$^3$ and molecular weight 284.18 g/mol is purchased from Sigma-Aldrich. BMIM-PF$_6$ is added to DIO in the weight proportion of 0.5:100 and stirred at 120 °C for 10 minutes. The concentration of ions in the resulting mixture, assuming full ionization of BMIM-PF$_6$, is $1.5 \times 10^{25}/m^3$.

Unidirectional alignment by polyimide PI2555. The aligning agent PI2555 and its solvent T9039, both purchased from HD MicroSystems, are combined in a 1:9 ratio. Glass substrates with patterned ITO electrodes are cleaned ultrasonically in distilled water and isopropyl alcohol, dried at 95ºC, cooled down to room temperature and blown with nitrogen. An inert N$_2$ environment is maintained inside the spin coater. Spin coating with the solution of the aligning agent is performed according to the following scheme: 1 sec @ 500 rpm → 30 sec @ 1500 rpm →1 sec @ 50 rpm. After the spin coating, the sample is baked at 95°C for 5 min, followed by 60 minutes of baking at 275°C. The spin coating produced the PI-2555 alignment layer with a thickness of 50 nm.

The PI2555 layer is buffed unidirectionally using a Rayon YA-19-R rubbing cloth (Yoshikawa Chemical Company, Ltd, Japan) of a thickness of 1.8 mm and filament density of 280/mm$^2$ to achieve a homogeneous planar alignment. An aluminum brick of a length of 25.5 cm, width of 10.4 cm, height of 1.8 cm, and weight of 1.3 kg, covered with the rubbing cloth, imposes a pressure of 490 Pa at a substrate and is moved ten times with a speed of 5 cm/s over the substrate; the rubbing length is about 1 m. Unidirectional rubbing of polyimide-coated substrates aligns the ferroelectric nematic liquid crystal (N$_F$) in a planar fashion. The polar character of surface alignment of spontaneous electric polarization **P** is evidenced by the response to the in-plane electric field in a planar cell confined between two PI2555 plates with parallel assembly of rubbing directions. These field experiments also suggest that the pretilt of **P** at the PI2555 substrates is negligibly small since the optical retardance Γ of the uniform domains in the presence of the in-plane electric field applied parallel to **P** and without it are close to each other and equal $\Delta nh$.

The PI2555 plates contain a pair of parallel transparent indium tin oxide (ITO) stripe electrodes separated along the rubbing direction **R** by a distance $l$ = 5-10 mm. A Siglent SDG1032X waveform generator and an amplifier (Krohn-Hite corporation) are used to apply an in-plane dc electric field $\mathbf{E} = E(0, \pm 1, 0)$. Since the cell thickness $h$ is much smaller than $l$, the electric field in this region is predominantly horizontal and uniform.

Degenerate alignment with polystyrene (PS). Glass substrates are cleaned ultrasonically in distilled water and isopropyl alcohol, dried at 95 ºC, cooled down to room temperature, and blown with nitrogen. Spin coating with the 1 % solution of PS in chloroform is performed for 30 s at 4000 rpm. After the spin coating, the sample is baked at 45 °C for 60 minutes. The thin part of the cell is glued with the epoxy glue Norland Optical Adhesive (NOA) 68 without spacers, the thick part is glued with the soda lime solid glass microspheres (Cospheric LLC.) of a diameter 53 µm mixed with an epoxy glue NOA 68. The dihedral angle of the empty cells is measured by an interference method at the wavelength 532 nm using a color filter (Thorlabs, Inc.) with 1 nm bandwidth.

Degenerate alignment with glycerin. A thin layer of glycerin (Fisher Scientific, CAS No. 56-81-5 with assay percent range 99-100 %w/v and density 1.261 g/cm$^3$ at 20 °C) is deposited onto the glass plates cleaned as described above. The height of the glycerin film is estimated to be $(500 \pm 100)$ nm by measuring the weight of the substrates before and after the deposition of glycerin and measuring the area of glycerin spreading.

Degenerate alignment at a free surface. A piece of crystallized DIO is placed onto the rubbed PI2555 at room temperature, heated to 120 °C, and cooled to the desired temperature at a rate of 5 °C/min. In the N, SmZ$_\text{A}$, and N$_\text{F}$ phases, DIO spreads over the surface and forms a film of ~10 μm thickness defined by the deposited mass and the film area. For example, $M = 1.4$ mg of DIO of a density $\rho = 1.35$ g/cm$^3$ at 60 °C (17), spread over an area $A = 1$ cm$^2$, produces a film of a thickness $h = \frac{M}{\rho A} \approx 10$ μm.

Cell assembly. The glass plates are assembled into flat cells of thickness defined by glass spacers. Wedge cells of a thickness varying from 0 to 53 μm and a small dihedral angle of 0.3° are prepared from PI2555 and PS plates. The rubbing direction **R** at the PI2555 layer is perpendicular to the thickness gradient to avoid twists caused by geometrical anchoring. The cell gap $h$ is measured by an interferometric technique recording transmission spectra of the light of wavelength between 400 nm and 900 nm, using a UV/VIS spectrometer Lambda 18 (Perkin Elmer). The thickness variations (caused by variations of the thickness of glass, coatings, and glue) over the area of the cell, approximately 1 cm$^2$, are about 0.2 μm or less. For example, typical thickness values measured in different $xy$ locations of a flat cell with glass spherical spacers of a diameter of 5 μm range from 4.82 μm to 5.00 μm.

Refractive indices measurements. The ordinary and extraordinary refractive indices are determined using the wedge cell technique and presented in Fig. S6. The wedge cell is formed by two unidirectionally rubbed PI2555-coated glass plates assembled in a parallel fashion such that the rubbing direction **R** is perpendicular to the thickness gradient. The thicknesses of the thinner and thicker parts of the wedge cells were set, respectively, employing the NOA 68 glue without spacers and the NOA 68 glue with premixed silica spacers. The dihedral angle $\alpha$ of an empty and filled cell is measured using the optical interference method. The interference fringes under the polarizing optical microscope are captured using a green interferometric filter with a center wavelength $\lambda = 532$ nm and a bandwidth of 1 nm. To determine $\alpha$, half of the wedge cell toward the thinner edge was filled with a liquid crystal using capillary action in the isotropic phase, while the other half toward the thicker edge was left empty and employed to capture the interference fringes. A 6-8% rise in $\alpha$ is observed after the cell is filled. As a result, the filled cell's wedge angle is used to increase the accuracy of measurement of the refractive indices. The temperature dependence of the wedge angle is also determined.

The reflection mode of the microscope is used. The cell is oriented so that the director parallel to the impinging polarization of light for measuring $n_e$ and perpendicular to it for $n_o$. In the wedge cell, the homogeneous region without domain walls was observed at the thinner edge, $h \leq 4$ μm, while the poly-domains were observed towards the thicker edge, for $h > 4$ μm. The homogeneous areas at $h \leq 4$ μm are used to calculate the $n_o$ and $n_e$ as $n_{o,e} = \frac{l\lambda}{2\alpha(S_{m+l}^{o,e} - S_m^{o,e})}$, where $l$ is the interference order, $S_{m+l}^{o,e} - S_m^{o,e}$ is the distance between the $m$ and $m + 1$ interference maxima.

Optical textures are recorded using a polarizing optical microscope Nikon Optiphot-2 with a QImaging camera and Olympus BX51 with an Amscope camera. PolScope MicroImager (Hinds Instruments) is used to map the director patterns and measure the optical retardance. The intensity of transmitted light in the textures is determined by using ImageJ software.

Numerical simulations use the experimentally determined ordinary and extraordinary refractive indices measured for the corresponding temperatures in DIO and RM734, Fig. S6. We use $n_o = 1.43$ and $n_e = 1.62$ for DIO at 60 °C and $n_o = 1.49$ and $n_e = 1.74$ for RM734 at 128 °C. The cell (with its midplane at $z = 0$ and with thickness $h$) is divided into thin slabs with thickness $\Delta z = h/N$, stacked along the $z$ axis. Each slab $i$ has a uniform optical axis oriented at an angle $\phi_i$ in the $xy$-plane. This angle varies linearly with the location $z_i$ of the slab mid-plane: $\phi_i = \tau(z_i + \frac{h}{2})/h$, with $z_i = -\frac{h}{2} + \frac{\Delta z}{2}, -\frac{h}{2} + \frac{3\Delta z}{2}, ..., \frac{h}{2} - \frac{\Delta z}{2}$. Each slab has an associated Jones matrix $J_i$ that takes into account both the rotation of the optical axis and the anisotropic optical response along the ordinary and extraordinary directions. The matrix reads $J_i = \begin{bmatrix} e^{i\delta_2}\sin^2\phi_i + e^{i\delta_1}\cos^2\phi_i & (e^{i\delta_1} - e^{i\delta_2})\sin\phi_i\cos\phi_i \\ (e^{i\delta_1} - e^{i\delta_2})\sin\phi_i\cos\phi_i & e^{i\delta_1}\sin^2\phi_i + e^{i\delta_2}\cos^2\phi_i \end{bmatrix}$, where $\delta_{1,2} = \frac{2\pi\Delta z}{\lambda}\, n_{o,e}$ are the phase shifts associated with ordinary and extraordinary directions, respectively. These matrices are all multiplied together to yield the full Jones matrix for the cell: $J = \prod_i J_i$, with the first element of the product corresponding to the uppermost slab. The polarizer is taken to be on the bottom of the cell and the analyzer is on top. The transmitted intensity, then, it given by

$$I = |\boldsymbol{v}_A^T J \boldsymbol{v}_P|^2,$$

where $\boldsymbol{v}_{A,P} = \begin{bmatrix} \cos\theta_{A,P} \\ \sin\theta_{A,P} \end{bmatrix}$ are the Jones vectors associated with the linear polarization directions of the analyzer and polarizer, respectively. The $T$ superscript indicates a row vector. The implementation of this calculation is done in Mathematica, using the following code:

```
(* Azimuthal angle ϕ *)
```

$\phi[\text{z\_}] := ((\tau) * \text{Pi}/180) * (z + \text{hh}/2)/\text{hh};$

```
(* Polar angle θ *)
```

$\theta[\text{z\_}] := 0.;$

```
(* Retardance factors for λλ=532 nm, cell thickness hh, and thickness of a single slice
dz=0.01-0.05 micrometers, refractive indices ne and no*)
```

$\delta\delta 1[\text{z\_}] := \frac{2*\pi*\text{dz}}{\lambda\lambda} * \left( \frac{\text{no}*\text{ne}}{\text{Sqrt}[\text{ne}^2*(\text{Sin}[\theta[z]])^2 + \text{no}^2*(\text{Cos}[\theta[z]])^2]} \right);$

$\delta\delta 2 = \frac{2*\pi*\text{dz}}{\lambda\lambda} * \text{no}$

```
(* Define the Jones matrix of a thin slab*)
```

$\text{slabMM}[\text{z\_}] := \{\{e^{i\delta\delta 2}\text{Sin}[\phi[z]]^2 + e^{i\delta\delta 1[z]}\text{Cos}[\phi[z]]^2, (e^{i\delta\delta 1[z]}$
$\qquad - e^{i\delta\delta 2})\text{Cos}[\phi[z]]\text{Sin}[\phi[z]]\}, \{(e^{i\delta\delta 1[z]}$
$\qquad - e^{i\delta\delta 2})\text{Cos}[\phi[z]]\text{Sin}[\phi[z]], e^{i\delta\delta 1[z]}\text{Sin}[\phi[z]]^2 + e^{i\delta\delta 2}\text{Cos}[\phi[z]]^2\}\}$

```
(* Calculate transmitted light intensity with varying angle between the polarizer and analyzer
*)
```

$\text{Intensity} = \{ \ \}; \text{Do}[\{\text{data} = \text{Reverse}[\text{Table}[\text{slabMM}[z], \{z, \frac{-\text{hh}}{2} + \frac{\text{dz}}{2}, \frac{\text{hh}}{2} - \frac{\text{dz}}{2}, \text{dz}\}]];$

$\text{MM} = \text{Dot}@@\text{data}; \text{Intensity}$
$\qquad = \text{Append}[\text{Intensity}, \{xx * 180/\text{Pi}, 100$
$\qquad * (\text{Abs}[\{\text{Cos}[xx], \text{Sin}[xx]\}.\text{MM}.\{\text{Cos}[\gamma], \text{Sin}[\gamma]\}])^2\}]; \}, \{xx, 0, \pi, 0.05\}]$

```
(* Plot the simulated transmittance vs. the uncrossing angle of the polarizers *)
```

$\text{Jones} = \text{ListPlot}[\text{Intensity}]$

**Fig. S1.**

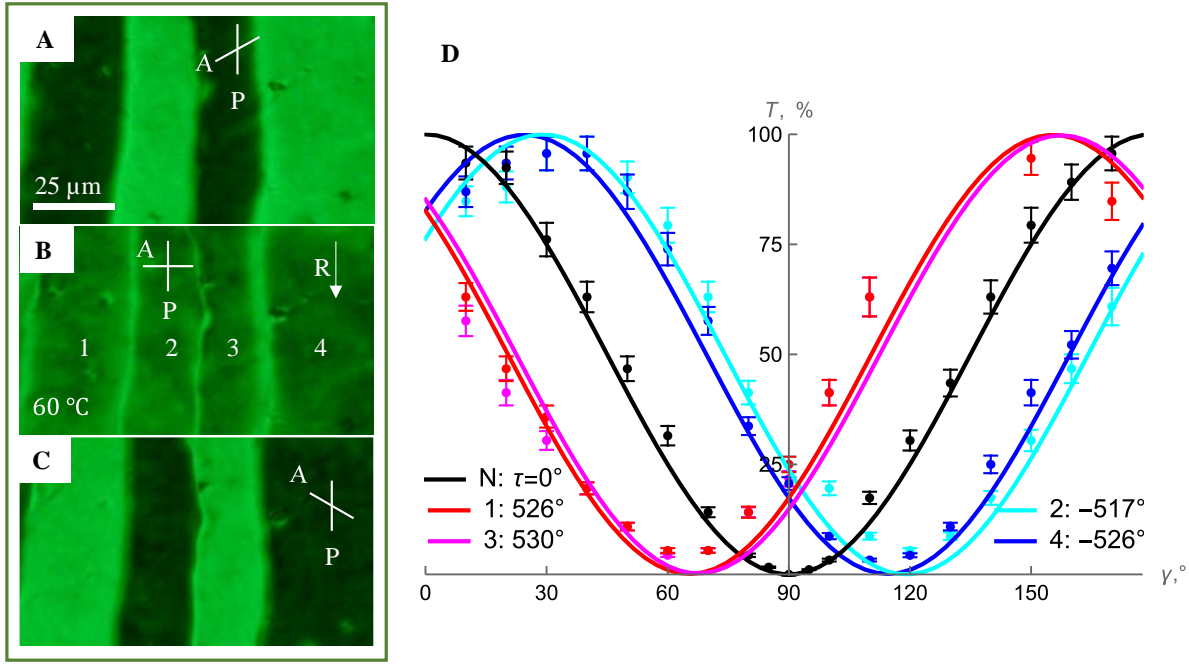

**Chiral ground states of DIO $N_F$ flat cell confined between PI2555 and glycerin-coated plate.** (**A, C**) Polarizing microscope textures for uncrossed polarizer and analyzer; note complementary contrast of the domains; temperature 60 °C; (**B**) the textures of the same domains for crossed polarizer and analyzer; (**D**) transmitted light intensity $T$ through four different $N_F$ domains labelled 1, 2, 3, and 4 in (B) as a function of the uncrossing angle $\gamma$; $T(\gamma)$ data are also shown for the N phase; solid curves are fits of the experimental data by numerical simulations based on Jones matrices with the twist angle $\tau$ as the fitting parameter shown for each domain. Cell thickness $h = 8$ μm. **R** is the rubbing direction at the bottom PI2555 substrate, A and P lines show the planes of polarization of the analyzer and polarizer, respectively.

**Fig. S2.**

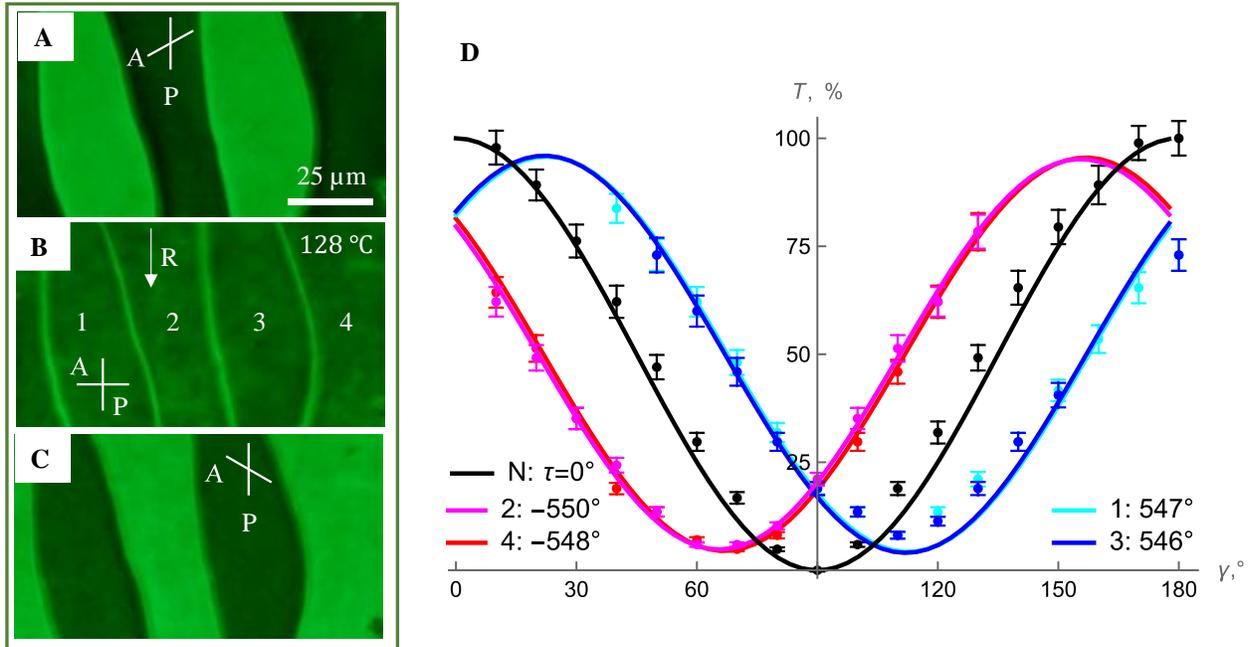

**Chiral ground states of RM734 N$_F$ flat cell confined between PI2555 and glycerin-coated plate.** (**A,C**) Polarizing microscope textures recorded with uncrossed polarizer and analyzer; note complementary contrast of the domains; temperature 128 °C; (**B**) the textures of the same domains for crossed polarizer and analyzer; (**D**) transmitted light intensity $T$ through four different N$_F$ domains labeled 1, 2, 3, and 4 in (**B**) as a function of the uncrossing angle $\gamma$; $T(\gamma)$ data are also shown for the N phase of RM734; solid curves are fits of the experimental data by numerical simulations based on Jones matrices with the twist angle $\tau$ as the fitting parameter shown for each domain. Cell thickness $h$ = 8 μm.

**Fig. S3.**

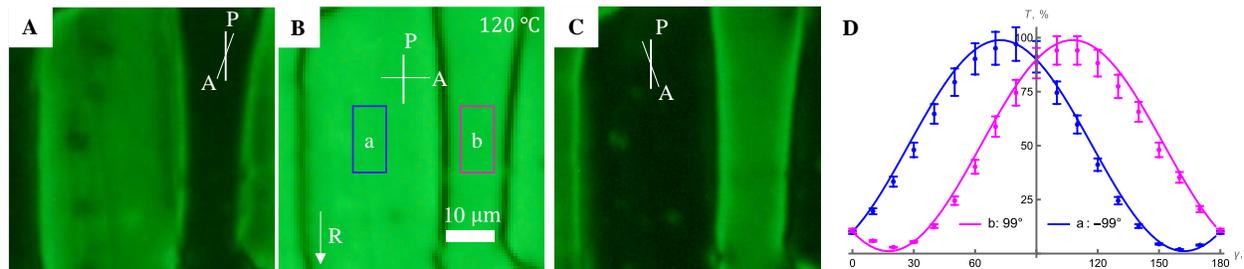

**Chiral ground states of RM734 N$_F$ flat cell confined between PI2555 and PS-coated plates.**
(**A,C**) Polarizing microscope textures recorded with uncrossed polarizer and analyzer; note complementary contrast of the domains; temperature 120 °C; (**B**) the textures of the same domains for crossed polarizer and analyzer; (**D**) transmitted light intensity $T$ through two N$_F$ domains labeled a and b in (B) as a function of the uncrossing angle $\gamma$; solid curves are fits of the experimental data by numerical simulations based on Jones matrices with the twist angle $\tau$ as the fitting parameter shown for each domain. Cell thickness $h = 3.5$ µm.

**Fig. S4.**

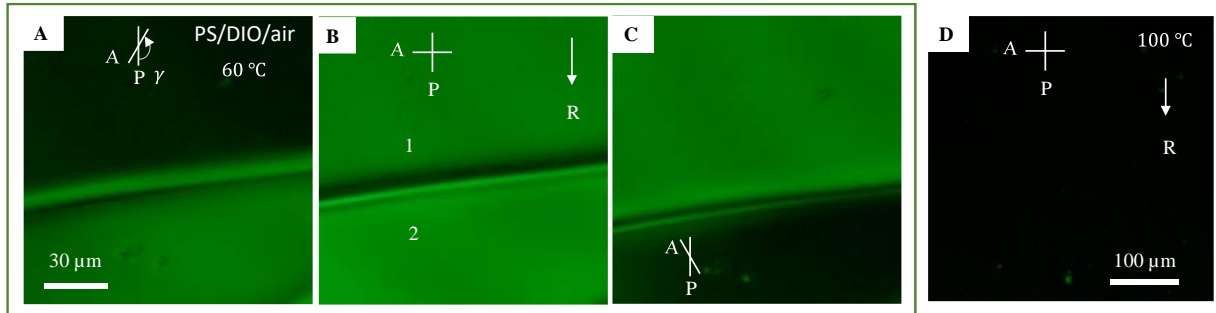

**Chiral ground states of DIO $N_F$ slab with free surface, placed on PI2555 plate.** (**A, C**) Polarizing microscope textures for uncrossed polarizer and analyzer; note complementary contrast of the two domains; (**B**) the textures of the same domains for crossed polarizer and analyzer; (**D**) the extinct texture of the same region in the N phase of DIO. Film thickness $h = 10$ μm. **R** is the rubbing direction at the bottom PI2555 substrate.

**Fig. S5.**

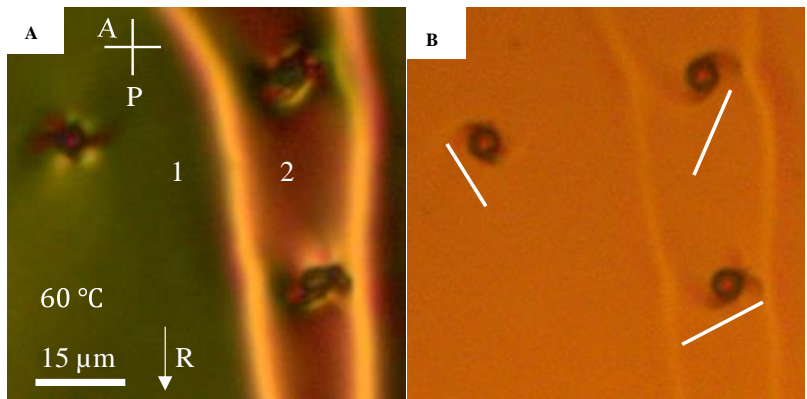

**Chiral ground states of DIO N$_F$ cell confined between PI2555 and PS plates visualized by spherical spacers.** (**A**) Polarizing microscope textures with crossed polarizer and analyzer of two twisted domains 1 and 2; domain 1 contains one silica sphere, domain 2 contains two spheres; (**B**) the same, unpolarized light; the axes of bipolar structures of the director around the spacers tilt in opposite directions in domains 1 and 2; white lines underneath the spheres show the approximate orientation of the axes. Cell thickness $h = 6$ µm; silica spheres of diameter 5 µm.

**Fig. S6.**

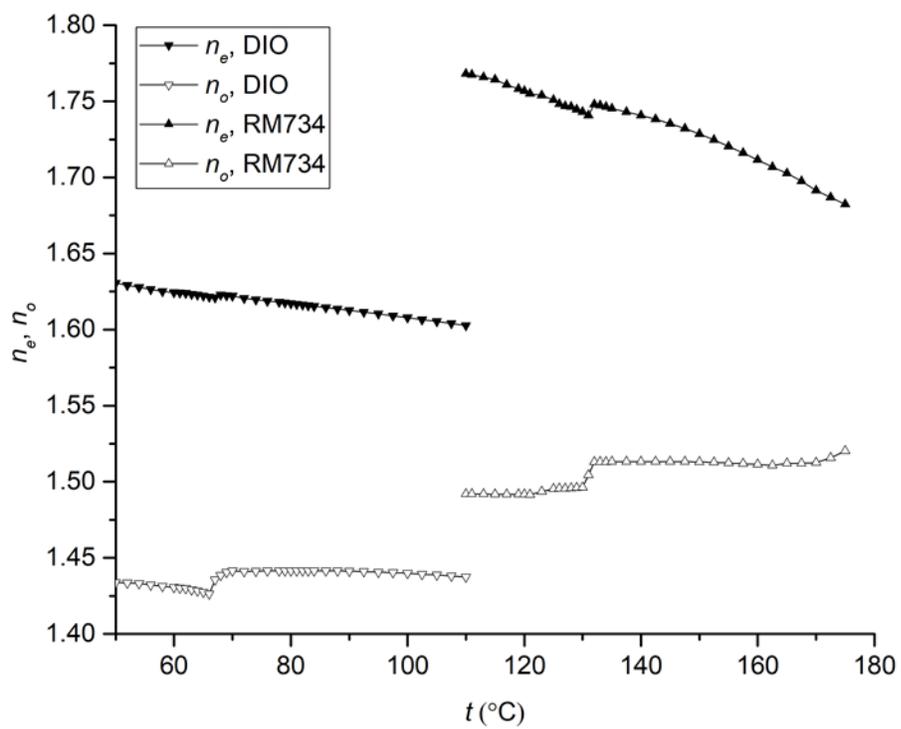

**Temperature dependencies of refractive indices of DIO and RM734 measured at 532 nm wavelength of light.**

**Fig. S7.**

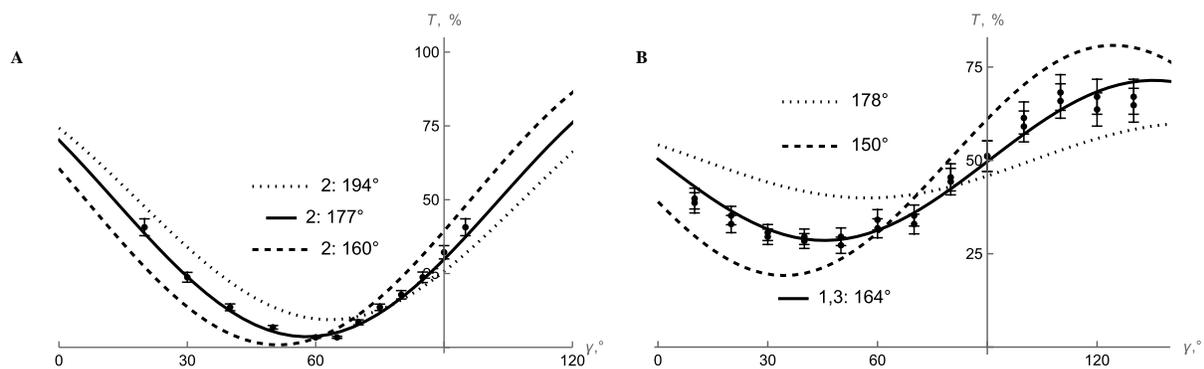

**Examples of fitting light transmission dependencies on the uncrossing angle.** (**A**) The best fit $\tau_2 = 177°$ for the DIO cell in Fig.1E is compared to fits with $\pm 10\% \tau_2$; (**B**) The best fit $\tau_{1,3} = 164°$ for the DIO cell in Fig.1J is compared to fits with $\pm 9\% \tau_{1,3}$.

**Fig. S8.**

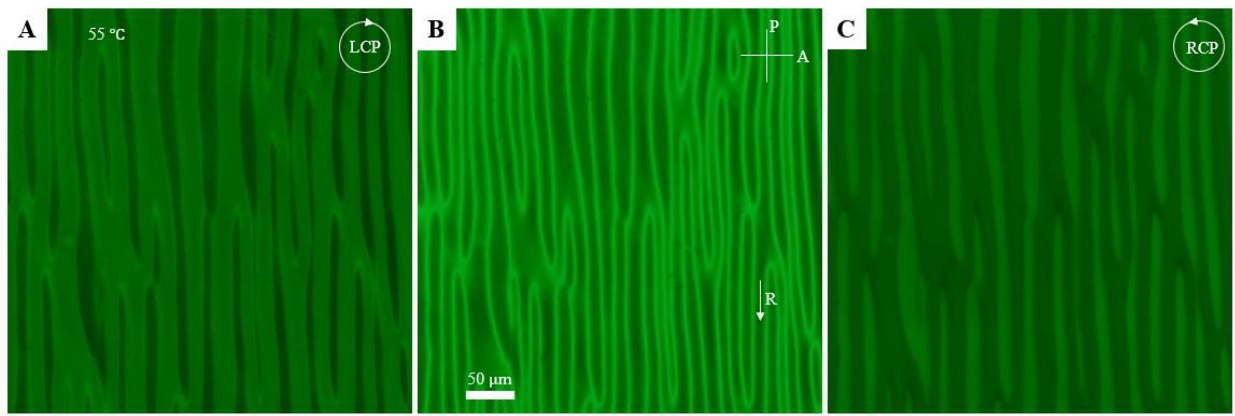

**Chiral ground states of DIO $N_F$ cells confined between PI2555 and PS observed under an optical microscope with circularly polarized light.** Textures of the twisted domains in **(A)** left-handed polarized light, **(B)** linearly polarized light, and **(C)** right-handed polarized light. Cell thickness 5 µm. Note the quasi one-dimensional periodicity of the domain structure.

**Fig. S9.**

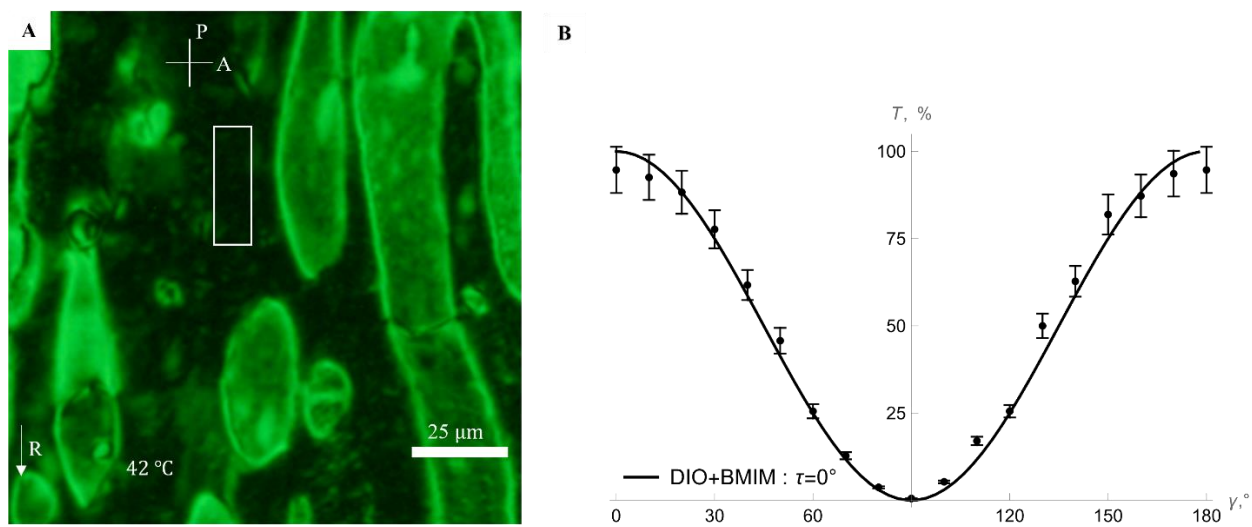

**Partially unwound ground state of N$_F$ flat cell of DIO doped with 0.5 wt. % of the ionic fluid.** (**A**) Polarizing microscope texture for crossed polarizers; bright domains with twist alternate with dark regions of a vanishing twist; (**B**) transmitted light intensity vs. $\gamma$ measured within the white rectangle in part (A) illustrating a vanishing twist angle. The N$_F$ is confined between a rubbed PI2555 plate and a glycerin plate, which sets azimuthally degenerate tangential anchoring. Cell gap $h = 8$ μm. The texture graininess might be caused by partial phase separation of the ionic fluid.

**Fig. S10.**

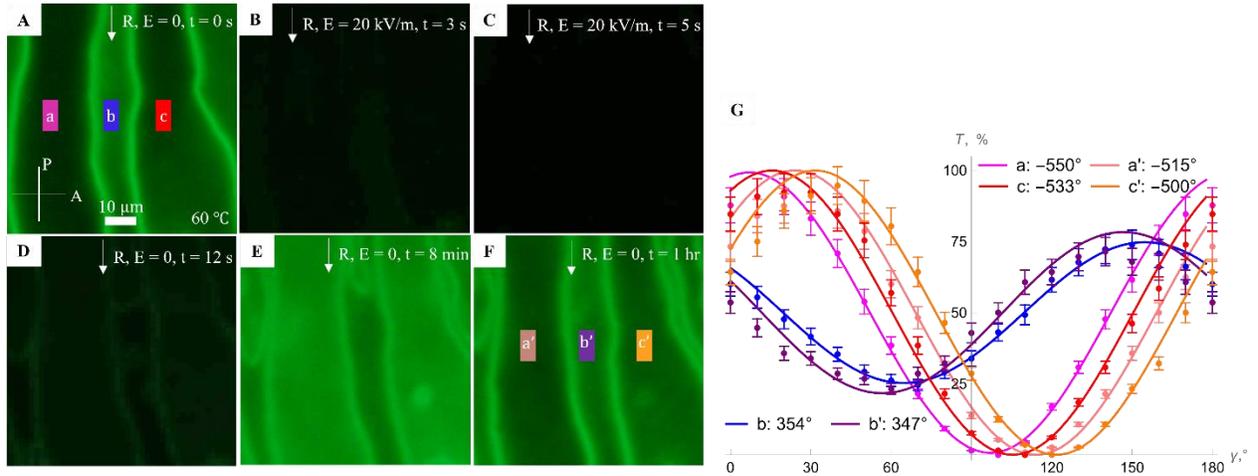

**Chiral ground state of DIO N$_F$ cell between PI2555 and PS plates removed by the electric field.** (**A**) Polarizing microscope textures with crossed polarizer and analyzer of twisted domains; (**B**) the same, at a moment of time $t = 3$ s, after the electric field 20 kV/m was applied as shown by a white arrow at $t_{on} = 1.5$ s; (**C**) the domains transform into a uniformly aligned homogeneous texture, which is extinct between the crossed polarizers; $t = 5$ s. (**D**) the twisted domains reappear after the electric field is switched off at $t_{off} = 5.2$ s. (**E**, **F**) the same re-appeared twisted domains at $t = 8$ min and at $t = 1$ hr, respectively. (**G**) transmitted light intensity $T$ through N$_F$ domains a, b, c in (A) and for the N$_F$ domains a', b', c' in (F) as a function of the angle $\gamma$ between the polarizer and analyzer; solid curves are fits by numerical simulations with the twist angle $\tau$ as the fitting parameter. Cell thickness $h = 7.8$ μm. **R** is the rubbing direction at the bottom PI2555 substrate, A and P lines show the planes of polarization of the analyzer and polarizer, respectively.